\documentclass[11pt]{article}
\usepackage[margin=1in]{geometry}
\usepackage{amsmath,amssymb,amsthm,mathtools}

\newtheorem{theorem}{Theorem}[section]
\newtheorem{lemma}[theorem]{Lemma}
\newtheorem{proposition}[theorem]{Proposition}
\newtheorem{corollary}[theorem]{Corollary}
\theoremstyle{definition}
\newtheorem{definition}[theorem]{Definition}

\newcommand{\one}{\mathbf{1}}
\newcommand{\R}{\mathbb{R}}
\newcommand{\E}{\mathbb{E}}

\newcommand{\rank}{\operatorname{rank}}

\newcommand{\polylog}{\operatorname{polylog}}
\newcommand{\low}{\mathrm{low}}
\newcommand{\high}{\mathrm{high}}

\title{Graded Projection Recursion (GPR): Corrections, Obstructions,\\ 
         and Conservative Approximate Matrix Multiplication}
\author{Jeffrey Uhlmann\\
Department of Electrical Engineering and Computer Science\\
University of Missouri-Columbia}
\date{}

\begin{document}
\maketitle

\begin{abstract}
Earlier versions proposed Graded Projection Recursion (GPR) as a deterministic packed-recursion framework for model-honest near-quadratic dense matrix multiplication. This revised version withdraws the exact dense matrix multiplication theorem and the downstream consequences that depended on it with a conservative AMM framework. The local ingredients remain useful as local tools: the three-band packing identity, scaled middle-band extraction under certified gaps, centering and reconstruction identities, and row/column normalization bounds. The gap in the earlier argument is global: the proof relied on a bounded active-state realization that would remove first-mismatch terms through the recursion. For arbitrary dense inputs this would require an exact equality filter over the inner index. We formulate this obstruction as a target-slice/equality-filter problem and give a rank/capacity argument against the natural separable active-state realization. The positive replacement is a conservative approximate matrix multiplication framework. For chosen protected left and right query subspaces, the low/marginal part of AB is computed exactly and an unbiased AMM primitive is applied only to the high/high residual. The resulting estimator is unbiased, preserves protected queries exactly in every realization, localizes stochastic error to the residual subspace, and inherits residual output-norm or query-risk guarantees from the underlying estimator.
\end{abstract}

%%%%%%%%%%%%%%%%%%%%%%%%%%%%%%%%%%%%%%
%%%%%%%%%%%%%%%%%%%%%%%%%%%%%%%%%%%%%%
%%%%%%%%%%%%%%%%%%%%%%%%%%%%%%%%%%%%%%
\section{Scope and replacement map}
\label{sec:replacement-map}

This manuscript is a corrected continuation of the GPR research program. The earlier versions studied recursive algebraic 
packing for dense matrix multiplication. The central local identity is:
\begin{equation}
 \label{eq:three-band-local-intro}
     (A_0+\beta^{-1}A_1)(B_0+\beta B_1) ~=~ \beta A_0B_1 + (A_0B_0+A_1B_1)+\beta^{-1}A_1B_0.
\end{equation}
The intended exact mechanism was to recursively extract the middle band while using centering and row/column normalization 
to control bit growth and extraction gaps. The local identity and the local extraction lemmas are correct under their stated 
hypotheses. The unsupported step is the global exact realization: the earlier argument relied on eliminating non-matched leaf 
terms at their first mismatch while maintaining only bounded active state. For arbitrary dense inputs, that active-band realization 
would have to implement an exact equality filter over the inner index.

The purpose of this version is to make the status of the old claims and the new claims unambiguous. Each exact claim below is 
withdrawn as an exact claim. The right-hand column records the narrower statement, with explicit hypotheses, that now occupies 
the same conceptual role in the corrected manuscript. Here ``replacement'' means that the AMM statement is a narrower 
statement asserted in place of the earlier exact statement, not a proof that the earlier exact theorem can be repaired.

%%%%%%%%%%%%%%%%%%%%%%%%%%%%%%%%%%%%%%
\subsection{Claim-by-claim replacement map}

\begin{center}
\begin{footnotesize}
\begin{tabular}{|p{0.25\linewidth}|p{0.7\linewidth}|}
\hline
{\bf Earlier exact claim} & {\bf Corrected replacement claim asserted here} \\ \hline
Exact near-quadratic deterministic dense matrix multiplication for arbitrary dense inputs. & Conservative AMM for arbitrary dense 
inputs: for chosen protected subspaces $U, V$, compute the low/marginal product component exactly and apply an unbiased 
AMM primitive only to the high/high residual $Q_U AB Q_V$. The output is unbiased, exact on all protected queries, and stochastic 
only on $U^\perp \times V^\perp$. \\ \hline
Matrix multiplication exponent $2$ or a soft-quadratic exact dense-MM kernel. & Near-quadratic conservative AMM contracts 
when the protected dimensions and residual sketch budget are polylogarithmic and the residual AMM primitive is near-quadratic. 
This is an approximate, stochastic, protected-query statement, not an exponent claim. \\ \hline

Recursive first-mismatch elimination with bounded exact active state. & Equality-filter obstruction plus residual localization: the 
exact active-band realization is not supplied, so the corrected construction makes chosen low/marginal directions exact and 
confines all random error to the unresolved high/high residual. \\ \hline

Exact near-quadratic downstream reductions for LUP, LDL, QR, GP regression, interior-point kernels, semiring tasks, and related 
applications. & Conditional task-level accuracy contracts: a downstream application may use the conservative estimator when its 
protected observables or aggregate/marginal statistics are covered by the chosen low spaces and the residual error tolerance is 
explicit. No exact downstream complexity consequence is claimed. \\ \hline

Bitwise equivalence to a floating-point GEMM comparator. & Finite-format claims require a separately specified comparator, 
representation, drift ledger, staging policy, and fallback condition. In a structured conservative representation, protected 
information can be exact relative to the represented algebra, then after dense finite-format materialization, leakage is a 
separate rounding-ledger issue. \\ \hline

Universal improvement over existing AMM methods. & Output-norm-compatible structural wrapper: when a specified unbiased 
output-norm AMM primitive is used on the residual, the familiar output-norm benchmark is inherited on the deflated residual 
$\|Q_U AB Q_V\|_F$ and protected queries are exact. This is a stronger structural accuracy contract than the corresponding 
norm-only statement when the protected-query or deflated-residual conditions in Section~\ref{sec:contract-comparison} apply, 
but it is not a universal improvement in every standard norm/runtime tradeoff. \\ \hline
\end{tabular}
\end{footnotesize}
\end{center}

%%%%%%%%%%%%%%%%%%%%%%%%%%%%%%%%%%%%%%
\subsection{Positive replacement claims}

The corrected manuscript makes the following replacement claims:
\begin{enumerate}
\item {\bf Local mechanisms}: The local three-band identity, scaled sieve under a certified gap, centering/reconstruction identities, 
         double-centering identities, and row(A)-column(B) Cauchy-Schwarz envelope are valid local tools.

\item {\bf Obstruction}: The earlier exact dense-MM proof lacks an executable bounded active-band realization. In the natural 
         separable model, exact realization of the inner-index equality filter requires linear lane capacity.

\item {\bf Layer 1 conservative AMM}: For protected subspaces $U,V$, the estimator:
          \begin{equation*}
               \widehat C ~=~ C^{\low}_{U,V}+Q_U S(Q_UA,BQ_V)Q_V
          \end{equation*}
          is unbiased, exact on $U$-left and $V$-right queries in every realization, localizes stochastic error to $U^\perp\times V^\perp$, 
          and inherits output-norm guarantees on the deflated residual.

\item {\bf Dyadic aggregate exactness}: For dyadic block-constant protected spaces $L_d$, all block row/column marginals and all 
          dyadic block aggregates through depth $d$ are exact in every realization.
  
\item {\bf Layer 2 query-risk refinement}: Given a paired-query risk, residual lanes can be sampled with probabilities proportional 
          to their query-risk norms, which gives the optimal independent one-lane Horvitz-Thompson law for the chosen residual risk, i.e.,
          singular query seminorms are handled by quotient or regularized full-support variants. This is an algorithmic near-quadratic 
          claim only when the required lane scores can be computed or certified within the stated work budget.

\item {\bf Near-quadratic regime}: If the protected dimensions, residual sketch budget, and any score-computation overhead is 
          $O(\polylog(n))$, and the residual AMM primitive has $O(n^2\polylog n)$ work, then the conservative estimator has 
          $O(n^2\polylog n)$ complexity.
\end{enumerate}

The below obstruction results do not prove that exact near-quadratic dense matrix multiplication is impossible. They identify 
the missing primitive in the earlier GPR proof and rule out the natural bounded separable active-state realization of that primitive. 
A future exact algorithm would have to use a genuinely different nonseparable target-slice mechanism, a structured equality-code 
algebra with sublinear channel cost, or a different computational target such as certified finite-format comparator equivalence.

%%%%%%%%%%%%%%%%%%%%%%%%%%%%%%%%%%%%%%
\subsection{Outline}
The manuscript is ordered to avoid ambiguity about the correction. Section~\ref{sec:replacement-map} gives the replacement 
map first. Section~\ref{sec:local-mechanisms} records local mechanisms that remain valid. Sections~\ref{sec:obstruction} 
and~\ref{sec:equality-filter} explain why those mechanisms do not establish exact arbitrary-dense near-quadratic multiplication. 
Sections~\ref{sec:conservative-amm} and~\ref{sec:query-risk} give the positive conservative-AMM and query-risk results. 
Section~\ref{sec:task-level-contracts} states how downstream uses should be reformulated as task-level contracts. 
Section~\ref{sec:relationship-standard-amm} compares the replacement with standard AMM, and Section~\ref{sec:limitations} 
states limitations and future directions.

%%%%%%%%%%%%%%%%%%%%%%%%%%%%%%%%%%%%%%
%%%%%%%%%%%%%%%%%%%%%%%%%%%%%%%%%%%%%%
%%%%%%%%%%%%%%%%%%%%%%%%%%%%%%%%%%%%%%
\section{Local GPR mechanisms that remain valid}
\label{sec:local-mechanisms}

This section states local ingredients that survive the correction. They should be treated as local algebraic and numerical tools, 
not as a complete exact dense matrix multiplication algorithm.

%%%%%%%%%%%%%%%%%%%%%%%%%%%%%%%%%%%%%%
\subsection{Three-band packing}

Let $A$ and $B$ be conformally partitioned into $2\times 2$ blocks. For a fixed output quadrant $(i,j)$ and a base $\beta>1$, 
define:
\begin{eqnarray}
     X_{ij} &=& A_{i1}+\beta^{-1}A_{i2},\\
     Y_{ij} &=& B_{1j}+\beta B_{2j},
\end{eqnarray}
then:
\begin{equation}
     X_{ij}Y_{ij} ~=~ \beta G_{ij}+T_{ij}+\beta^{-1}L_{ij},
\end{equation}
where:
\begin{eqnarray}
     T_{ij} &=& A_{i1}B_{1j}+A_{i2}B_{2j}=(AB)_{ij},\\
     G_{ij} &=& A_{i1}B_{2j},\\
     L_{ij} &=& A_{i2}B_{1j}.
\end{eqnarray}
This identity is purely algebraic and is not the source of the flaw.

%%%%%%%%%%%%%%%%%%%%%%%%%%%%%%%%%%%%%%
\subsection{Scaled two-round sieve}

Let the local coefficient lattice be:
\begin{equation}
     L_{\mathrm{coeff}}(\sigma) ~=~ D^{-2}2^{-\sigma}\mathbb{Z},
\end{equation}
with lift $M=D^2 2^\sigma$. Define:
\begin{eqnarray}
  R_\sigma(z)             &=& M^{-1}\lfloor Mz\rceil,\\
  S_{\beta,\sigma}(z) &=& R_\sigma(z)-\beta R_\sigma(z/\beta),
\end{eqnarray}
where $\lfloor\cdot\rceil$ denotes nearest-integer rounding with a fixed deterministic tie-breaking convention.

\begin{lemma}[Local scaled-sieve exactness]
Suppose
\begin{equation}
     z ~=~ \beta U+T+\frac{L}{\beta},
\end{equation}
where $U,T\in L_{\mathrm{coeff}}(\sigma)$ and $L\in\R$. If, for some $\delta>0$,
\begin{equation}
\label{eq:scaled-gap}
     M\left(\frac{|L|}{\beta}+\frac{|T|}{\beta}+\frac{|L|}{\beta^2}\right) ~\leq~ \frac{1}{2}-\delta,
\end{equation}
then:
\begin{equation}
     S_{\beta,\sigma}(z) ~=~ T.
\end{equation}
\end{lemma}
\begin{proof}
Multiplying by $M$ gives:
\begin{equation}
     Mz ~=~ \beta(MU)+MT+\frac{ML}{\beta},
\end{equation}
where $MU$ and $MT$ are integers. The gap condition implies that $Mz/\beta$ rounds to $MU$ and $Mz$ rounds to 
$\beta(MU)+MT$. Subtracting $\beta$ times the former rounded value from the latter gives $MT$, and rescaling gives $T$.
\end{proof}
The important qualification is that this lemma is local. It says that if a node presents a genuine three-band quantity with a 
certified gap, then the middle band is recovered. It does not by itself show that a recursive computation can maintain only 
bounded exact active-band state for arbitrary dense inputs.

%%%%%%%%%%%%%%%%%%%%%%%%%%%%%%%%%%%%%%
\subsection{Centering and reconstruction}

Let $\alpha,\rho\in\R^n$ and define:
\begin{eqnarray}
     A_0 &=& A-\one\alpha^T,\\
     B_0 &=& B-\rho\one^T,
\end{eqnarray}
then:
\begin{equation}
\label{eq:centering-identity}
     AB ~=~ A_0B_0+\one(\alpha^T B_0)+(A_0\rho)\one^T+(\alpha^T\rho)\one\one^T.
\end{equation}
Thus a centered product can be converted back to the original product by quadratic-time matrix-vector and outer-product 
bookkeeping once $A_0B_0$ is available.

%%%%%%%%%%%%%%%%%%%%%%%%%%%%%%%%%%%%%%
\subsection{Row(A)-column(B) normalization}

\begin{lemma}[Dimension-free dot-product envelope]
Let $A',B'\in\R^{m\times m}$ satisfy:
\begin{eqnarray}
     \|A'_{i,:}\|_2 &\leq& 1 ~\text{for all}~i,\\
     \|B'_{:,j}\|_2 &\leq& 1 ~\text{for all}~j,
\end{eqnarray}
then: 
\begin{equation}
     |(A'B')_{ij}| ~\leq~ 1 ~\text{for every}~ i,j.
\end{equation}
\end{lemma}
\begin{proof}
$(A'B')_{ij}=\langle A'_{i,:},B'_{:,j}\rangle$, so the result follows from Cauchy-Schwarz.
\end{proof}
This is the legitimate role of row(A)-column(B) normalization: it supplies a magnitude envelope for local bands. 
It does not compress information and does not itself give a global exact fast multiplication algorithm.

%%%%%%%%%%%%%%%%%%%%%%%%%%%%%%%%%%%%%%
\subsection{Double centering as conservation}

Let $u=n^{-1/2}\one$ and $P=I-uu^T$. For any product $C=XY$:
\begin{equation}
\label{eq:double-centering-decomposition}
     C ~=~ PCP+PCuu^T+uu^TCP+uu^TCuu^T.
\end{equation}
The non-centered part can be computed without forming $C$:
\begin{eqnarray}
     Cu &=& X(Yu),\\
     u^TC &=& (u^T X)Y,\\
     u^TCu &=& (u^T X)(Yu).
\end{eqnarray}
If $\widetilde C$ is an unbiased estimator of $C$, then:
\begin{equation}
     \widehat C ~=~ (C-PCP)+P\widetilde C P
\end{equation}
is also unbiased, and:
\begin{equation}
     \widehat C-C ~=~ P(\widetilde C-C)P,
\end{equation}
so row and column sums of the error vanish exactly. This observation is the block-constant special case of the 
conservative framework in Section~\ref{sec:conservative-amm}.

%%%%%%%%%%%%%%%%%%%%%%%%%%%%%%%%%%%%%%
\subsection{Double centering is not exact compression}

The preceding subsection explains why double centering is useful for conservation and approximate computation. 
It is also important to record what double centering does not do. It does not reduce the exact worst-case 
information content of the centered core.

\begin{proposition}[Full centered residual family]
Let $P=I-uu^T$ and let:
\begin{equation}
     \mathcal M_P ~=~ \{X\in\R^{n\times n}:~ X ~=~ PXP\}
\end{equation}
be the doubly centered matrix space. Define the low/marginal parts:
\begin{eqnarray}
     A_{\leq} &=& A-PAP,\\
     B_{\leq} &=& B-PBP,
\end{eqnarray}
and consider the mean-peeled predictor:
\begin{equation}
\label{eq:mean-peeled-predictor}
     \widetilde C_{\mathrm{dc}}(A,B) ~=~ A_{\leq}B_{\leq}+A_{\leq}(PBP)+(PAP)B_{\leq}.
\end{equation}
Equivalently, by expanding $A=A_{\leq}+PAP$ and $B=B_{\leq}+PBP$:
\begin{equation}
     \widetilde C_{\mathrm{dc}}(A,B) ~=~ AB-(PAP)(PBP).
\end{equation}
Thus the predictor is not defined by first forming $AB$, i.e., Equation~\ref{eq:mean-peeled-predictor} is the computable 
low/marginal expansion, and its residual family is exactly $\mathcal M_P$:
\begin{equation}
     \{AB-\widetilde C_{\mathrm{dc}}(A,B):~ A,B\in\R^{n\times n}\} ~=~ \mathcal M_P
\end{equation}
and has dimension $(n-1)^2$.
\end{proposition}
\begin{proof}
The residual is $(PAP)(PBP)$ for every $A,B$, so it is doubly centered and lies in $\mathcal M_P$. 
Conversely, let $C\in\mathcal M_P$. Choose $A=C$ and $B=P$, then $PAP=C$ and $PBP=P$, so the residual is 
$CP=C$. Thus every doubly centered matrix occurs as a residual. Because $P$ has rank $n-1$, the space 
$\mathcal M_P$ is isomorphic to $\R^{(n-1)\times(n-1)}$ and has dimension $(n-1)^2$.
\end{proof}

This proposition is a guardrail for the corrected manuscript. Centering can isolate low and marginal components, and it can 
make those components exact in a conservative estimator. But if the remaining centered core is required exactly for arbitrary 
dense inputs, it is still an arbitrary dense product on an $(n-1)$-dimensional subspace. The conservative framework below 
changes the target by approximating that residual. It does not claim that double centering alone gives an exact 
low-dimensional correction.

%%%%%%%%%%%%%%%%%%%%%%%%%%%%%%%%%%%%%%
%%%%%%%%%%%%%%%%%%%%%%%%%%%%%%%%%%%%%%
%%%%%%%%%%%%%%%%%%%%%%%%%%%%%%%%%%%%%%
\section{Why the exact deterministic theorem is not established}
\label{sec:obstruction}

The local identity and the local sieve are not enough. The unsupported step in the earlier exact theorem is global: the proof 
needs an executable active-band realization that removes first-mismatch terms while preserving all fully matched inner-index 
contributions.

%%%%%%%%%%%%%%%%%%%%%%%%%%%%%%%%%%%%%%
\subsection{The active-band invariant gap}

One possible induction invariant is:
\begin{equation}
\label{eq:full-product-invariant}
     \mathrm{KernelMM}(X,Y) ~=~ XY.
\end{equation}
Under this invariant, recursive products of packed operands are full numeric products, including all off-band material inherited 
from ancestors. This gives clean local extraction statements, but it does not by itself prevent off-band material from reappearing 
as deeper numeric content. To control exact active state requires an invariant of the form:
\begin{equation}
\label{eq:active-band-invariant}
     \mathrm{BandMM}_E(X,Y) ~=~ \Pi_E(XY),
\end{equation}
where $\Pi_E$ projects onto a desired degree window or target slice. This is a different computational object. 
It requires a concrete realization of the projection without carrying a large coefficient table. The earlier exact proof 
moved between \eqref{eq:full-product-invariant} and \eqref{eq:active-band-invariant} without establishing such a realization.

%%%%%%%%%%%%%%%%%%%%%%%%%%%%%%%%%%%%%%
\subsection{First-mismatch intuition}

A leaf term in the recursive expansion corresponds to a path label for the left inner index and a path label for the right inner index. 
The needed terms are those for which the two labels agree at every bit, whereas a nonmatched term has a first node at which the 
bit choices differ. Locally, that term is routed into an off-band at that node. This intuition is valid, but the executable realization is
problematic. More specifically, to make first-mismatch elimination exact with bounded active state, the algorithm must implement 
an equality filter over all inner-index paths.

%%%%%%%%%%%%%%%%%%%%%%%%%%%%%%%%%%%%%%
\subsection{A capacity obstruction in the separable active-state model}

The following elementary rank argument captures the obstruction relevant to the earlier proof strategy:

\begin{theorem}[Diagonal-contraction capacity]
Let $F$ be a field with finite-dimensional vector spaces $V_L,V_R,W$ over $F$, and let:
\begin{equation}
     \mu:V_L\times V_R\to W
\end{equation}
be bilinear and $\varphi:W\to F$ be linear. If there exist $u_1,\ldots,u_N\in V_L$ and $v_1,\ldots,v_N\in V_R$ such that:
\begin{equation}
  \varphi(\mu(u_p,v_q)) ~=~\delta_{pq},~~ 1\leq p,q\leq N,
\end{equation}
then $\dim V_L\geq N$ and $\dim V_R\geq N$.
\end{theorem}
\begin{proof}
Define $M\in F^{N\times N}$ by $M_{pq}=\varphi(\mu(u_p,v_q))$, then by hypothesis, $M=I_N$, so $\rank(M)=N$. 
For fixed $u\in V_L$, the map:
\begin{equation}
     v\mapsto \varphi(\mu(u,v))
\end{equation}
is a linear functional on $V_R$ that depends linearly on $u$. Thus the row space spanned by the functionals associated 
with $u_1,\ldots,u_N$ has dimension at most $\dim V_L$, and $\dim V_L\geq N$ because its realized matrix on $v_1,\ldots,v_N$ 
has rank $N$. The argument with left and right interchanged gives $\dim V_R\geq N$.
\end{proof}
\begin{corollary}[Bounded separable active state is insufficient]
At recursion depth $L=\log_2 n$, exact first-mismatch elimination for arbitrary dense $n\times n$ multiplication requires an 
$n$-lane equality filter over the inner index. In the natural separable bilinear active-state model formalized by the theorem, 
exact realization of that filter requires lane capacity $\Omega(n)$. In particular, polylogarithmic exact active state is not 
enough in that model.
\end{corollary}
The corollary is not a lower bound for all possible algorithms for matrix multiplication. It is a diagnosis of the missing step 
in the earlier GPR proof and of a broad family of local separable active-band implementations. A successful exact continuation 
would have to evade this model by using a genuinely nonseparable target-slice primitive or by changing the computational 
target.

%%%%%%%%%%%%%%%%%%%%%%%%%%%%%%%%%%%%%%
\subsection{Why this does not contradict the local lemmas}

There is no contradiction between local sieve exactness and the global obstruction. The local lemma says that if a packed 
product already has the form:
\begin{equation}
     z ~=~ \beta U+T+\frac{L}{\beta}
\end{equation}
with $U,T$ on the local lattice and a certified gap, then the middle band $T$ is recovered. The global obstruction says that 
arranging recursively that all non-target contributions have already been removed, while keeping only polylogarithmic exact 
state for arbitrary dense inner-index paths, requires an equality filter that cannot be represented in bounded separable state.

%%%%%%%%%%%%%%%%%%%%%%%%%%%%%%%%%%%%%%
\section{Equality filtering and target slices}
\label{sec:equality-filter}

The obstruction is clearest when matrix multiplication is written as equality filtering over the inner index.
Specifically, for $A,B\in\R^{n\times n}$:
\begin{equation}
\label{eq:equality-filter}
     C_{ij} ~=~ \sum_{p,q=1}^n A_{ip}B_{qj}\delta_{pq}.
\end{equation}
Thus exact dense matrix multiplication applies the equality kernel $\delta_{pq}$ to separated left and right inner-index data.
Equivalently, introduce a Laurent variable $t$ and define:
\begin{eqnarray}
     f_i(t) &=&\sum_{p=1}^n A_{ip}t^p,\\
     g_j(t) &=& \sum_{q=1}^n B_{qj}t^{-q}.
\end{eqnarray}
Then:
\begin{equation}
     C_{ij} ~=~ [t^0]f_i(t)g_j(t).
\end{equation}
Matrix multiplication is therefore a target-slice product: compute the zero slice in the inner-index grading for every output 
pair $(i,j)$.

The local GPR identity is a one-bit equality filter. For a one-bit split, the terms $A_0B_0$ and $A_1B_1$ correspond to equal-bit 
choices and appear in the middle band. The terms $A_0B_1$ and $A_1B_0$ correspond to mismatches and appear in off-bands. 
The earlier exact proof attempted to compose this local filter recursively without carrying all active matched lanes. The capacity 
obstruction explains why the local filter is not enough.

Standard global equality codes, such as Fourier or CRT idempotent representations of $\delta_{pq}$, do give exact equality tests, 
but they carry $\Theta(n)$ equality channels unless an additional nonseparable batching primitive is supplied. 
Scalar middle-product methods do not remove the difficulty by themselves, because the target coefficient is matrix-valued:
\begin{equation}
     [t^0]\sum_{p,q} A_{:,p}B_{q,:}t^{p-q} ~=~ \sum_p A_{:,p}B_{p,:}.
\end{equation}
Computing this coefficient is exactly the dense product.

The post-GPR exact target can be stated independently:
\begin{quote}
Given dense $A\in\R^{n\times N}$ and $B\in\R^{N\times n}$ with $N=\Theta(n)$, compute the zero slice:
\begin{equation}
  [t^0]\left(\sum_{p=1}^N A_{:,p}t^p\right)
        \left(\sum_{q=1}^N B_{q,:}t^{-q}\right) ~=~ AB
\end{equation}
without materializing all off-slice channels or all equality channels.
\end{quote}
No such exact target-slice primitive is established here.

%%%%%%%%%%%%%%%%%%%%%%%%%%%%%%%%%%%%%%
\section{Conservative approximate matrix multiplication}
\label{sec:conservative-amm}

This section gives the corrected positive framework. The principle is projection-first: identify the queries, marginals, 
or decision directions that must be exact, then compute all product interactions involving those directions exactly, and 
then approximate only the residual subspace.

%%%%%%%%%%%%%%%%%%%%%%%%%%%%%%%%%%%%%%
\subsection{Protected subspaces}

Let $U,V\subseteq\R^n$ be protected left and right subspaces, with orthogonal projectors $P_U,P_V$, and let:
\begin{eqnarray}
     Q_U &=& I-P_U,\\
     Q_V &=& I-P_V.
\end{eqnarray}
For $C=AB$, define the protected low/marginal component:
\begin{equation}
\label{eq:low-part}
     C^{\low}_{U,V} ~=~ P_UAB+ABP_V-P_UABP_V
\end{equation}
and the high/high residual:
\begin{equation}
\label{eq:high-residual}
     C^{\high}_{U,V} ~=~ Q_UABQ_V.
\end{equation}
Then:
\begin{equation}
     AB ~=~ C^{\low}_{U,V}+C^{\high}_{U,V}.
\end{equation}
The following elementary bookkeeping lemma is included to make the cost of the exact protected part explicit.

\begin{lemma}[Cost of the protected part]
Let $\dim U=k_U$ and $\dim V=k_V$, and suppose orthonormal basis matrices for $U$ and $V$ are given. 
The protected part $C^{\low}_{U,V}$ in \eqref{eq:low-part} can be computed without forming $AB$ using
$O(n^2(k_U+k_V)+nk_Uk_V)$ arithmetic operations, plus the overhead cost of materializing the chosen output 
representation.
\end{lemma}
\begin{proof}
Let $U_0\in\R^{n\times k_U}$ and $V_0\in\R^{n\times k_V}$ have orthonormal columns spanning $U$ and $V$. 
The complexity to compute $U_0^TA$ and $BV_0$ is $O(n^2k_U)$ and $O(n^2k_V)$, respectively, so:
\begin{eqnarray}
     P_UAB &=& U_0(U_0^TA)B,\\
     ABP_V &=& A(BV_0)V_0^T,
\end{eqnarray}
and the overlap is:
\begin{equation}
     P_UABP_V ~=~ U_0\bigl((U_0^TA)(BV_0)\bigr)V_0^T,
\end{equation}
where the middle $k_U\times k_V$ product costs $O(nk_Uk_V)$. The remaining multiplications by the skinny basis 
matrices are included in the displayed dense-output cost, which in a structured output representation may be stored 
factored.
\end{proof}
Note that the spaces $U$ and $V$ are part of the accuracy specification, so they may be fixed in advance, e.g., chosen from 
known task queries or selected from the inputs by a deterministic preprocessing step. If randomized pilot data is used 
to choose them, the production residual sketch should be conditionally unbiased after the selected spaces are fixed, e.g., 
by sample splitting. Reusing the same randomness to both select $U,V$ and to form the final residual estimator can 
introduce selection bias and would require a separate argument. 

\begin{lemma}[Protected queries]
If $E=Q_UEQ_V$, then $x^TEy=0$ whenever $x\in U$ or $y\in V$.
\end{lemma}
\begin{proof}
If $x\in U$, then $Q_Ux=0$, so $x^TEy=x^TQ_UEQ_Vy=0$. The case $y\in V$ is identical.
\end{proof}

%%%%%%%%%%%%%%%%%%%%%%%%%%%%%%%%%%%%%%
\subsection{The conservative wrapper}

Let $S_r(X,Y)$ be any unbiased approximate matrix multiplication primitive for $XY$, where $r$ denotes its sketch or 
sampling budget:
\begin{equation}
     \E S_r(X,Y) ~=~ XY.
\end{equation}

\begin{theorem}[Conservative AMM wrapper]
Let $A,B\in\R^{n\times n}$ and choose protected subspaces $U,V$. Define:
\begin{equation}
\label{eq:conservative-estimator}
  \widehat C ~=~ C^{\low}_{U,V}+Q_U S_r(Q_UA,BQ_V)Q_V.
\end{equation}
Then:
\begin{enumerate}
     \item $\E\widehat C=AB$.
     \item $\widehat C-AB=Q_U(\widehat C-AB)Q_V$ in every realization.
     \item Every query $x^T\widehat C y$ with $x\in U$ or $y\in V$ is exact in every realization.
     \item All stochastic error is confined to the high/high residual subspace $U^\perp\times V^\perp$.
\end{enumerate}
\end{theorem}
\begin{proof}
By unbiasedness of $S_r$:
\begin{eqnarray}
     \E[Q_US_r(Q_UA,BQ_V)Q_V] &=& Q_U(Q_UABQ_V)Q_V\\
                                                    &=& Q_UABQ_V,
\end{eqnarray}
and adding $C^{\low}_{U,V}$ gives $\E\widehat C=AB$. Finally:
\begin{equation}
     \widehat C-AB ~=~ Q_U\left(S_r(Q_UA,BQ_V)-Q_UABQ_V\right)Q_V,
\end{equation}
which proves high/high localization. Exactness of protected queries follows from the preceding lemma.
\end{proof}

\begin{proposition}[Whole-product conservative projection]
Let $\widetilde C$ be any unbiased estimator of $C=AB$, and define:
\begin{equation}
     \widehat C_{\mathrm{proj}} ~=~ C^{\low}_{U,V}+Q_U\widetilde C Q_V.
\end{equation}
Then $\E\widehat C_{\mathrm{proj}}=C$, all $U$-left and $V$-right queries are exact, and:
\begin{equation}
     \widehat C_{\mathrm{proj}}-C ~=~ Q_U(\widetilde C-C)Q_V.
\end{equation}
Consequently, for every unitarily invariant norm $\|\cdot\|_{\mathrm{ui}}$:
\begin{equation}
     \|\widehat C_{\mathrm{proj}}-C\|_{\mathrm{ui}}
                       ~\leq~ \|\widetilde C-C\|_{\mathrm{ui}}
\end{equation}
in every realization.
\end{proposition}
\begin{proof}
The identity $C=C^{\low}_{U,V}+Q_UCQ_V$ gives the displayed error formula, and unbiasedness follows by expectation. 
Exact protected queries follow from high/high localization. Orthogonal left and right projections are contractions for all 
unitarily invariant norms.
\end{proof}

%%%%%%%%%%%%%%%%%%%%%%%%%%%%%%%%%%%%%%
\subsection{Minimum-change interpretation}

The conservative correction is also the least-change way to impose the declared exact low/marginal information on a 
raw estimator. This helps separate the method from an ad hoc postprocessing rule.

\begin{proposition}[Frobenius-nearest constrained correction]
Let $C=AB$ and let $\widetilde C$ be any raw product estimate and define the affine constraint set:
\begin{equation}
     \mathcal A_{U,V}(C) ~=~\{X\in\R^{n\times n}:~ P_UX ~=~ P_UC,~~ XP_V ~=~ CP_V\},
\end{equation}
then:
\begin{equation}
     C^{\low}_{U,V}+Q_U\widetilde C Q_V ~=~ \operatorname*{argmin}_{X\in\mathcal A_{U,V}(C)}\|X-\widetilde C\|_F .
\end{equation}
\end{proposition}
\begin{proof}
Every $X\in\mathcal A_{U,V}(C)$ has the unique decomposition:
\begin{equation}
     X ~=~ C^{\low}_{U,V}+Q_UXQ_V.
\end{equation}
The orthogonal block decomposition of $Z=Q_UXQ_V$ gives:
\begin{equation}
     X-\widetilde C ~=~ \bigl(C^{\low}_{U,V}-P_U\widetilde C-\widetilde C P_V+P_U\widetilde C P_V\bigr)
                                       + Q_U(Z-\widetilde C)Q_V,
\end{equation}
where the two summands lie in orthogonal Frobenius subspaces. The first summand is fixed by the constraints, and the second 
is minimized uniquely by choosing $Z=Q_U\widetilde C Q_V$.
\end{proof}

\subsection*{Rao-Blackwell analogy}
The protected marginals are deterministic information about $C=AB$. Projecting a raw unbiased estimate onto the affine 
space with those known marginals preserves unbiasedness and removes stochastic degrees of freedom in the projected-out 
channels. This is analogous to a Rao-Blackwell improvement, although no probabilistic conditioning formalism is needed for 
the algebraic claim above.

%%%%%%%%%%%%%%%%%%%%%%%%%%%%%%%%%%%%%%
\subsection{Output-norm residual guarantee}

\begin{definition}[Unbiased output-norm primitive]
A randomized AMM primitive $S_r$ has output-norm rate $\alpha(n,r)$ and work $W(n,r)$ if, for all square inputs 
$X,Y\in\R^{n\times n}$:
\begin{equation}
  \E S_r(X,Y) ~=~ XY,~~ \E\|S_r(X,Y)-XY\|_F^2 ~\leq~ \alpha(n,r)\|XY\|_F^2,
\end{equation}
and its arithmetic work is at most $W(n,r)$ (rectangular variants may be substituted when available).
\end{definition}

\begin{corollary}[Deflated output-norm guarantee]
If $S_r$ has output-norm rate $\alpha(n,r)$, then the conservative estimator \eqref{eq:conservative-estimator} satisfies:
\begin{eqnarray}
     \E\|\widehat C-AB\|_F^2 &\leq& \alpha(n,r)\|Q_UABQ_V\|_F^2\\
                                               &\leq& \alpha(n,r)\|AB\|_F^2.
\end{eqnarray}
If $k_U+k_V$ and the residual sketch budget are polylogarithmic, and $W(n,r)=O(n^2\polylog n)$ at that budget, then the 
wrapper is near-quadratic.
\end{corollary}
\begin{proof}
Apply the output-norm guarantee to $X=Q_UA$ and $Y=BQ_V$, so $XY=Q_UABQ_V$. The final inequality follows because 
multiplication on the left and right by orthogonal projectors is Frobenius-contracting.
\end{proof}

\begin{definition}[Deflation ratio]
For $AB\neq 0$, define:
\begin{equation}
  \rho_{U,V}(A,B) ~=~ \frac{\|Q_UABQ_V\|_F^2}{\|AB\|_F^2}\in[0,1].
\end{equation}
For a fixed wrapped output-norm primitive, the Frobenius benchmark is multiplied by the deflation ratio $\rho_{U,V}(A,B)$. 
If $\rho_{U,V}(A,B)=1$, no Frobenius-constant improvement is implied, although exact protected-query guarantees may still 
be valuable. If $\rho_{U,V}(A,B)<1$, the deflated residual benchmark is strictly smaller for that input pair.
\end{definition}

%%%%%%%%%%%%%%%%%%%%%%%%%%%%%%%%%%%%%%
\subsection{Adaptive choice of protected subspaces}
The protected spaces may be chosen from the input matrices, but the production residual estimator must remain unbiased 
after the choice is fixed. A clean sufficient condition is conditional unbiasedness.

\begin{lemma}[Conditional protected-space selection]
Let $U,V$ be random protected subspaces chosen from $A,B$ and auxiliary randomness $\zeta$. Suppose $\zeta$ is 
independent of the production sketch randomness and, conditionally on the realized $U,V$:
\begin{equation}
     \E \left[S_r(Q_UA,BQ_V)\mid U,V\right] ~=~ Q_UABQ_V,
\end{equation}
then the conservative estimator is unbiased unconditionally and has exact $U$-left and $V$-right queries for the realized 
protected spaces in every realization.
\end{lemma}
\begin{proof}
Condition on $U,V$. The fixed-subspace conservative wrapper gives conditional expectation $AB$ and high/high localization 
relative to the realized projectors. Taking expectation over $\zeta$ gives unconditional unbiasedness. The protected-query 
identities are deterministic consequences of high/high localization and therefore hold for each realized pair of subspaces.
\end{proof}

Note that the conditional unbiasedness hypothesis may fail if the same random sketch is used both to choose $U,V$ and to 
estimate the residual. A clean implementation should use deterministic protected spaces, independent pilot sketches, sample 
splitting, or a cross-fitting argument with explicit independence bookkeeping.

%%%%%%%%%%%%%%%%%%%%%%%%%%%%%%%%%%%%%%
\subsection{Dyadic aggregate specialization}

Let $\mathcal D_d$ be the dyadic partition of $\{1,\ldots,n\}$ into $2^d$ blocks. Let $L_d$ be the subspace of vectors that are 
constant on each block, and set $U=V=L_d$. For $I\in\mathcal D_d$, define $u_I=|I|^{-1/2}\one_I$. For an error matrix $E$, 
define the block-aggregate error at level $\ell$ by:
\begin{equation}
     \mathrm{BAE}_\ell(E)^2 ~=~ \frac{1}{|\mathcal D_\ell|^2}
                                                       \sum_{I,J\in\mathcal D_\ell}(u_I^TEu_J)^2.
\end{equation}

\begin{corollary}[Exact aggregate accuracy]
For the conservative estimator with $U=V=L_d$:
\begin{equation}
  \mathrm{BAE}_\ell(\widehat C-AB) ~=~ 0,~~ (0\leq \ell\leq d),
\end{equation}
in every realization. Equivalently, all dyadic block row/column aggregate queries through depth $d$ are exact.
\end{corollary}
\begin{proof}
For $\ell\leq d$, every normalized depth-$\ell$ block indicator lies in $L_d$. The conservative error is $Q_{L_d}EQ_{L_d}$, 
so every such aggregate query has zero error.
\end{proof}

%%%%%%%%%%%%%%%%%%%%%%%%%%%%%%%%%%%%%%
\subsection{Finite-format caveat}

The exact protected-query statements are algebraic statements about the represented estimator. In exact arithmetic, staged 
arithmetic, or a structured representation that stores the exact protected part separately from the residual, the protected queries 
are exact in the represented algebra. If the final output is materialized as a fully rounded dense array, then exact preservation 
can be degraded by the final rounding step. That degradation is a representation error and must be accounted for by a separate 
finite-precision ledger, in the ordinary spirit of finite-precision error analysis \cite{Higham2002}. The conservative theorem is not 
a bitwise floating-point equivalence statement unless such a ledger and comparator model are explicitly supplied.

%%%%%%%%%%%%%%%%%%%%%%%%%%%%%%%%%%%%%%
%%%%%%%%%%%%%%%%%%%%%%%%%%%%%%%%%%%%%%
%%%%%%%%%%%%%%%%%%%%%%%%%%%%%%%%%%%%%%
\section{Query-risk residual sampling}
\label{sec:query-risk}

Frobenius error is not always the right way to state commensurate accuracy. The conservative wrapper removes all protected query 
mass exactly, so the remaining stochastic problem is to approximate the high/high residual in the directions that matter to the task.

Let $\mathcal Q$ be a probability distribution over paired queries $(x,y)\in\R^n\times\R^n$, and define the query-risk seminorm:
\begin{equation}
     \|E\|_{\mathcal Q}^2 ~=~ \E_{(x,y)\sim\mathcal Q}\left[(x^TEy)^2\right].
\end{equation}
Equivalently:
\begin{equation}
     \|E\|_{\mathcal Q}^2 ~=~ \langle E,R_{\mathcal Q}(E)\rangle_F, 
\end{equation}
where:
\begin{equation}
     R_{\mathcal Q}(E) ~=~ \E_{(x,y)\sim\mathcal Q}[xx^TEyy^T]
\end{equation}
is a positive semidefinite fourth-order risk operator. The seminorm may be singular, i.e., directions invisible to the query law have 
zero query risk.

Suppose the high/high residual has a lane decomposition:
\begin{equation}
     K_{U,V} ~=~ Q_UABQ_V=\sum_{a\in\mathcal A}Z_a.
\end{equation}
Choose probabilities $p_a>0$ on the nonzero lanes and draw $a_1,\ldots,a_s$ independently, and define:
\begin{eqnarray}
     \widehat K_{p,s} &=& \frac{1}{s}\sum_{t=1}^s \frac{Z_{a_t}}{p_{a_t}},\\
     \widehat C_{p,s} &=& C^{\low}_{U,V}+\widehat K_{p,s},
\end{eqnarray}
then $\E\widehat C_{p,s}=AB$ and the conservative high/high localization is preserved.

\begin{theorem}[Query-risk formula and optimal one-lane law]
For every full-support sampling law $p$:
\begin{equation}
\label{eq:query-risk-formula}
     \E\|\widehat C_{p,s}-AB\|_{\mathcal Q}^2 ~=~ \frac{1}{s}\left[\sum_{a\in\mathcal A}\frac{\|Z_a\|_{\mathcal Q}^2}{p_a}
                                                                                   - \|K_{U,V}\|_{\mathcal Q}^2\right].
\end{equation}
If $\|Z_a\|_{\mathcal Q}>0$ for every nonzero lane, the minimizing distribution among full-support independent 
one-lane samplers is:
\begin{equation}
\label{eq:query-optimal-law}
     p_a^* ~=~ \frac{\|Z_a\|_{\mathcal Q}}{\sum_b\|Z_b\|_{\mathcal Q}}.
\end{equation}
\end{theorem}
\begin{proof}
For one sample $V=Z_a/p_a$:
\begin{equation}
     \E V ~=~ \sum_a Z_a ~=~ K_{U,V}
\end{equation}
and:
\begin{eqnarray}
     \E\|V\|_{\mathcal Q}^2 &=& \sum_a p_a\frac{\|Z_a\|_{\mathcal Q}^2}{p_a^2}\\
                                            &=& \sum_a\frac{\|Z_a\|_{\mathcal Q}^2}{p_a}.
\end{eqnarray}
Averaging $s$ independent copies divides the variance by $s$, giving \eqref{eq:query-risk-formula}. The only term 
depending on $p$ is $\sum_a w_a^2/p_a$, where $w_a=\|Z_a\|_{\mathcal Q}$. Cauchy-Schwarz gives:
\begin{equation}
     \left(\sum_a w_a\right)^2 ~\leq~ \left(\sum_a\frac{w_a^2}{p_a}\right)\left(\sum_a p_a\right),
\end{equation}
with equality when $p_a\propto w_a$.
\end{proof}

\subsection*{Singular query risks}
If some nonzero lanes have $\|Z_a\|_{\mathcal Q}=0$, the visible-lane law gives the optimum for query-unbiasedness 
but may fail full matrix-unbiasedness. A full-support matrix-unbiased law can approach the visible-lane optimum by 
assigning small probability to query-invisible lanes, or by using the regularized weights:
\begin{equation}
     p_{a,\lambda} ~\propto~ \sqrt{\|Z_a\|_{\mathcal Q}^2+\lambda\|Z_a\|_F^2},~~ \lambda>0.
\end{equation}

%%%%%%%%%%%%%%%%%%%%%%%%%%%%%%%%%%%%%%
\subsection{Score-computation caveat}

The optimal sampling law \eqref{eq:query-optimal-law} is a statistical optimality statement. It becomes an algorithmic 
near-quadratic statement only when the lane weights $\|Z_a\|_{\mathcal Q}$ can be computed, estimated, or certified 
within the same budget. The following elementary empirical-query case gives one useful sufficient condition:
\begin{proposition}[Empirical query scores for residual column-row lanes]
Let $X=Q_UA$ and $Y=BQ_V$ so that the residual is:
\begin{eqnarray}
     K_{U,V} &=& XY ~=~ \sum_{r=1}^n Z_r,\\
     Z_r &=& X_{:,r}Y_{r,:}.
\end{eqnarray}
Let $\mathcal Q_M$ be the empirical paired-query law on query pairs $(x^{(m)},y^{(m)})$, $1\leq m\leq M$, then:
\begin{equation}
     \|Z_r\|_{\mathcal Q_M}^2 ~=~ \frac{1}{M}\sum_{m=1}^M \bigl((x^{(m)})^T X_{:,r}\bigr)^2 \bigl(Y_{r,:}y^{(m)}\bigr)^2.
\end{equation}
All $n$ scores can be computed in $O(Mn^2)$ arithmetic operations by forming the two thin products:
\begin{equation}
  [x^{(1)}  \cdots  x^{(M)}]^T X ~\text{and}~ Y [y^{(1)} \cdots y^{(M)}].
\end{equation}
\end{proposition}
\begin{proof}
For a rank-one lane $Z_r=X_{:,r}Y_{r,:}$:
\begin{equation}
     (x^{(m)})^T Z_r y^{(m)} ~=~ \bigl((x^{(m)})^T X_{:,r}\bigr)\bigl(Y_{r,:}y^{(m)}\bigr).
\end{equation}
Squaring and averaging over the empirical law gives the displayed score formula. The first thin product contains all 
values $(x^{(m)})^T X_{:,r}$, and the second contains all values $Y_{r,:}y^{(m)}$. Each costs $O(Mn^2)$ for dense inputs.
\end{proof}

\begin{corollary}[Near-quadratic empirical query-risk regime]
For the one-shot residual column-row sampler, suppose $\dim U+\dim V=\polylog(n)$, $M=\polylog(n)$ empirical paired 
queries are used for scoring, and $s=\polylog(n)$ residual lanes are sampled and materialized in the dense-output model. 
Then the exact protected part, score computation, and sampled residual output can be formed in $O(n^2\polylog n)$ 
arithmetic operations.
\end{corollary}
\begin{proof}
The exact protected part costs $O(n^2\polylog n)$ by the cost lemma in Section~\ref{sec:conservative-amm}. The empirical 
scores cost $O(Mn^2)=O(n^2\polylog n)$ by the proposition. Materializing $s$ rank-one residual lanes costs 
$O(sn^2)=O(n^2\polylog n)$. Summing the three terms gives the claim.
\end{proof}

%%%%%%%%%%%%%%%%%%%%%%%%%%%%%%%%%%%%%%
%%%%%%%%%%%%%%%%%%%%%%%%%%%%%%%%%%%%%%
%%%%%%%%%%%%%%%%%%%%%%%%%%%%%%%%%%%%%%
\section{Task-level contracts and downstream claims}
\label{sec:task-level-contracts}

The earlier versions stated downstream algorithmic consequences as if an exact GPR matrix multiplication primitive had been 
established. This manuscript does not make such reductions. The safe replacement is a task-level accuracy contract: specify 
which observables a downstream computation requires, protect those observables when possible, and bound the residual 
effect otherwise.

Let $\Phi=\{\phi_1,\ldots,\phi_m\}$ be a finite family of linear task statistics on the product matrix, and suppose a downstream 
decision or reported quantity depends on $\phi_j(AB)$. For bilinear statistics $\phi_j(E)=x_j^TEy_j$, conservative projection 
gives the exact identity:
\begin{equation}
     \phi_j(\widehat C-AB) ~=~ (Q_Ux_j)^T(\widehat C-AB)(Q_Vy_j),
\end{equation}
so the statistic is exact whenever $x_j\in U$ or $y_j\in V$.

\begin{proposition}[Decision invariance from protected or bounded task statistics]
Suppose a downstream decision is invariant whenever each statistic $\phi_j$ is perturbed by at most $\Delta_j>0$. 
If an estimator $\widehat C$ satisfies:
\begin{equation}
     |\phi_j(\widehat C-AB)|\leq \Delta_j,~~ j=1,\ldots,m,
\end{equation}
with probability at least $1-\eta$, then the downstream decision agrees with the exact-product decision with probability at 
least $1-\eta$. In particular, any statistic with a left argument in $U$ or a right argument in $V$ contributes zero stochastic 
drift in every realization.
\end{proposition}
\begin{proof}
This is just the definition of the decision slack. The deterministic zero-drift statement follows from high/high localization of 
the conservative error.
\end{proof}

Consequently, downstream statements in this version are conditional contracts, not complexity reductions: exactness is 
asserted only for protected observables. Unprotected effects require a stated norm, query-risk, concentration, finite-format 
ledger, or fallback theorem.

%%%%%%%%%%%%%%%%%%%%%%%%%%%%%%%%%%%%%%
%%%%%%%%%%%%%%%%%%%%%%%%%%%%%%%%%%%%%%
%%%%%%%%%%%%%%%%%%%%%%%%%%%%%%%%%%%%%%
\section{Relationship to standard AMM and to the earlier GPR program}
\label{sec:relationship-standard-amm}

The conservative framework is presented as a structural wrapper, not as a blanket replacement for existing AMM methods. 
Existing randomized AMM schemes already provide unbiased estimators and strong norm guarantees. Classical examples 
include randomized matrix sampling and compressed matrix multiplication \cite{DKM2006,Pagh2013}. Recent output-norm 
AMM work provides a useful benchmark class: for square inputs and a runtime parameter $r\leq n$, one such result gives 
an $O(n^2(r+\log n))$-time estimator whose unbiased variant has expected squared Frobenius error at most 
$(n/r)\|AB\|_F^2$ \cite{UffenheimerWeinstein2026}. This citation is used only to indicate the kind of primitive the wrapper 
can call, i.e., the contribution here is not a new survey or a blanket comparison theorem. The present contribution is a 
structural wrapper around such primitives: exact protected queries, high/high residual localization, and a query-risk calculus 
for the residual.

The conservative replacement claims are not a universal improvement over every AMM method in every standard norm. 
They confer an improvement only relative to a declared accuracy contract and cost model: the protected subspaces must be 
low-dimensional enough to compute within the target work budget, and the task or benchmark must either value protected 
queries or have nonzero low/marginal product content. Equivalently, in the partial order below, the wrapper is strictly 
stronger only when it adds exact task-relevant invariants not present in the comparison contract, or when the residual 
benchmark is genuinely deflated; otherwise it is a non-worsening structural wrapper, not a new standard-norm improvement 
claim.

%%%%%%%%%%%%%%%%%%%%%%%%%%%%%%%%%%%%%%
\subsection{A partial-order comparison of accuracy contracts}
\label{sec:contract-comparison}

The comparison used in this manuscript is a comparison of {\em contracts}, not a claim that one algorithm is uniformly 
better than every other algorithm under every metric.

\begin{definition}[Accuracy-contract comparison]
Fix a target product $C=AB$, a work scale $\mathcal W(n)$, a norm or seminorm $\mu$, and a family 
$\mathcal Q_{\mathrm{task}}$ of bilinear queries $x^TCy$ that the application declares important. A contract 
$\mathfrak C_1$ refines a contract $\mathfrak C_0$ for this task, written informally as 
$\mathfrak C_1\succeq \mathfrak C_0$, if:
\begin{enumerate}
     \item $\mathfrak C_1$ preserves the same unbiasedness or bias guarantee as $\mathfrak C_0$.
     \item The work of $\mathfrak C_1$ remains within the declared scale $\mathcal W(n)$.
     \item The stated $\mu$-error bound of $\mathfrak C_1$ is no larger than that of $\mathfrak C_0$.
     \item The queries guaranteed exact by $\mathfrak C_1$ contain the queries guaranteed exact by $\mathfrak C_0$, and 
               contain any protected queries from $\mathcal Q_{\mathrm{task}}$.
     \item Any stochastic-error localization in $\mathfrak C_1$ is an additional restriction on where error may occur, not a 
               replacement for the norm or query-risk bound.
\end{enumerate}
The refinement is strict for the declared contract if at least one of the comparison components is strict and none is worse. 
This is the sense in which the conservative wrapper is stronger than a norm-only AMM statement.
\end{definition}

\begin{proposition}[When the conservative wrapper improves a contract]
Let $C=AB$, choose protected subspaces $U,V$, and write:
\begin{eqnarray}
     C &=& C^{\low}_{U,V}+Q_UCQ_V,\\
     C^{\low}_{U,V} &=& P_UC+CP_V-P_UCP_V.
\end{eqnarray}
Assume the exact protected computation is within the declared work budget, i.e.:
\begin{equation}
     n^2(\dim U+\dim V)+n\dim U\dim V ~=~ O(\mathcal W(n)).
\end{equation}

\begin{enumerate}
\item {\bf Whole-product wrapping}:  If $\widetilde C$ is any unbiased estimator of $C$, then:
          \begin{equation}
               \widehat C ~=~ C^{\low}_{U,V}+Q_U\widetilde C Q_V
          \end{equation}
          is unbiased, has exact $U$-left and $V$-right queries, and satisfies:
          \begin{equation}
               \widehat C-C ~=~ Q_U(\widetilde C-C)Q_V.
          \end{equation}
          Therefore, for every unitarily invariant norm $\|\cdot\|_{\mathrm{ui}}$,
          \begin{equation}
               \|\widehat C-C\|_{\mathrm{ui}} ~\leq~ \|\widetilde C-C\|_{\mathrm{ui}}
          \end{equation}
          in every realization.

\item {\bf Residual output-norm wrapping}:  If $S_r$ is an unbiased output-norm primitive with rate $\alpha(n,r)$, 
          then the residual wrapper satisfies
          \begin{eqnarray}
               \E\|\widehat C-AB\|_F^2 &\leq& \alpha(n,r)\|Q_UABQ_V\|_F^2\\
                                                        &\leq& \alpha(n,r)\|AB\|_F^2.
          \end{eqnarray}
\end{enumerate}
Consequently, relative to the corresponding norm-only contract, the conservative wrapper gives a strict contract 
improvement whenever either: (i) the task declares some protected query or low/marginal invariant important that 
the comparison contract does not already guarantee exactly, or (ii) the low/marginal component $C^{\low}_{U,V}$ 
is nonzero, equivalently:
\begin{equation}
     \rho_{U,V}(A,B) ~:=~ \frac{\|Q_UABQ_V\|_F^2}{\|AB\|_F^2}<1 ~(AB\neq 0),
\end{equation}
so that the residual Frobenius benchmark is strictly smaller. If the protected dimensions are too large, if the task ignores 
protected queries and localization, and if $AB$ is entirely high/high with respect to $U,V$, then this proposition gives no 
strict improvement in the ordinary Frobenius benchmark.
\end{proposition}
\begin{proof}
The cost condition is the skinny-product cost from \eqref{eq:low-part}. For part~1, the identity:
\begin{equation}
     C ~=~ C^{\low}_{U,V}+Q_UCQ_V
\end{equation}
gives:
\begin{equation}
     \widehat C-C ~=~ Q_U(\widetilde C-C)Q_V.
\end{equation}
Unbiasedness follows by taking expectations, and exact protected queries follow from the protected-query lemma. 
Multiplication by orthogonal projectors is contractive for every unitarily invariant norm, proving the norm inequality.

Part~2 is the deflated output-norm corollary from Section~\ref{sec:conservative-amm}. Orthogonality of the four 
blocks $U\times V$, $U\times V^\perp$, $U^\perp\times V$, and $U^\perp\times V^\perp$ gives:
\begin{equation}
     \|C\|_F^2 ~=~ \|C^{\low}_{U,V}\|_F^2+\|Q_UCQ_V\|_F^2.
\end{equation}
Thus the residual Frobenius benchmark is strictly smaller exactly when $C^{\low}_{U,V}\neq 0$. The final statement is 
just the definition of strict contract refinement applied to the guarantees above.
\end{proof}

Note that the phrase ``stronger than a norm-only AMM contract'' is used only in the contract sense just defined, not 
as a blanket algorithmic ranking. Fix an unbiased output-norm primitive $S_r$ and wrap that same primitive by the 
conservative construction. The wrapped estimator has no worse Frobenius guarantee than applying the same rate to 
the full product, because:
\begin{equation}
     \|Q_UABQ_V\|_F ~\leq~ \|AB\|_F.
\end{equation}
It also has exact $U$-left and $V$-right queries in every realization, which an unwrapped norm-only estimator generally 
does not provide. It is not described here as a universal improvement over all possible AMM algorithms: pure high/high 
stress cases may have deflation ratio 1, and a comparison metric that ignores protected queries will not provide a 
structural gain.
\begin{center}
\begin{tabular}{|p{0.2\linewidth}|p{0.35\linewidth}|p{0.375\linewidth}|}
\hline
{\bf Method or viewpoint} & {\bf Typical role} & {\bf What the conservative wrapper adds} \\ \hline
Column-row residual sampling & Unbiased lane estimator & Exact protected queries and residual-only sampling \\ \hline
Sketch-based AMM & Black-box approximate product & Deflated residual target $Q_U AB Q_V$ \\ \hline
Output-norm AMM & Error benchmark based on product output & Same benchmark on the high/high residual, plus exact low queries \\ \hline
Finite-format GPR ideas & Decision invariance under margins & A clearer protected-subspace target and fallback boundary \\ \hline
\end{tabular}
\end{center}

This table is just a taxonomy of guarantees, not a ranking. The wrapper improves the declared accuracy contract, but it does 
not by itself establish a better runtime/error tradeoff in every standard benchmark. The strongest safe statement is that 
GPR-inspired conservative projection can be combined with suitable AMM primitives to obtain exact protected low/marginal 
behavior while approximating only the residual.

%%%%%%%%%%%%%%%%%%%%%%%%%%%%%%%%%%%%%%
%%%%%%%%%%%%%%%%%%%%%%%%%%%%%%%%%%%%%%
%%%%%%%%%%%%%%%%%%%%%%%%%%%%%%%%%%%%%%
\section{Limitations and future directions}
\label{sec:limitations}

The corrected framework separates three branches:
\begin{description}
\item[Exact arbitrary-dense multiplication] - The exact deterministic theorem from earlier versions is withdrawn. A future 
         exact route would need a nonseparable target-slice primitive, a structured equality-code algebra whose target extraction 
         avoids the $n$-channel contraction, or another mechanism outside the bounded separable active-state model discussed 
         above. No such route is established here.

\item[Conservative AMM]- The projection-first conservative AMM theorem is the current rigorous positive result. It gives exact 
         protected queries and residual stochastic guarantees for arbitrary dense inputs whenever the protected subspaces and 
         residual sketch budget are small enough for the desired complexity target.

\item[Finite-format or comparator-faithful computation] - The earlier drift-ledger and decision-invariance ideas may still be 
         useful for finite-format targets, but they require explicit comparator definitions, finite-precision ledgers, staged precision 
         or fallback policies, and a proof that the certified regime is non-vacuous. Such claims are not conflated here with exact 
         dense matrix multiplication.

\item[Downstream reductions] - The earlier downstream case studies are not reinstated as consequences of this manuscript. 
         A downstream use may be developed only after specifying its task observables, the protected subspaces that make those 
         observables exact, and the residual norm, query-risk, or finite-format ledger that controls the remaining error. In that sense, 
         future reductions are task-level accuracy contracts rather than automatic consequences of a fast exact matrix multiplication 
         primitive.
\end{description}

In summary, this version is a correction and replacement of the earlier exact dense matrix multiplication claim, not an attempt to 
restate that claim in approximate language. The local GPR mechanisms remain useful, but they do not establish exact 
near-quadratic arbitrary dense matrix multiplication. The corrected positive framework is conservative approximate matrix 
multiplication: exact protected low/marginal queries, stochastic error confined to a high/high residual, output-norm or query-risk 
guarantees inherited from the chosen residual estimator, and task-level contracts for any downstream use.

\end{document}